\documentstyle[multicol,aps,epsf,here]{revtex} 
\begin{document}
\draft

\title{Photoelectron Spectra of Aluminum Cluster Anions:
 Temperature Effects  and Ab-Initio Simulations  }

\author{Jaakko Akola$^1$, Matti Manninen$^1$,
Hannu H\"akkinen$^2$, Uzi Landman$^2$,
Xi Li$^3$, and Lai-Sheng Wang$^3$}
\address{$^1$Department of Physics, University of Jyv\"askyl\"a,
FIN-40351 Jyv\"askyl\"a, Finland}
\address{$^2$School of Physics, Georgia Institute of Technology, 
Atlanta, GA 30332-0430}
\address{$^3$Department of Physics, Washington State University,
 Richland, WA 99352, and W. R. Wiley Environmental Molecular 
Sciences Laboratory, Pacific Northwest National Laboratory, 
MS K8-88, P. O. Box 999, Richland, WA 99352}  
\date{May 24, 1999}
\maketitle

\begin{abstract}
 Photoelectron  (PES) spectra 
from aluminum cluster anions, Al$_n^-$ ($12\leq n\leq 15$),
at various temperature regimes,
were  studied  
 using  ab-initio molecular
dynamics simulations and
experimentally. 
The calculated PES spectra,
obtained via shifting of the 
simulated electronic densities of
states by the self-consistently
determined values of the asymptotic
exchange-correlation potential, agree
well with the measured ones, allowing
reliable structural assignments and
theoretical estimation of the
clusters' temperatures.

\end{abstract}

\begin{multicols}{2}
\narrowtext

Photoelectron spectroscopy (PES)
is a rich source of information
pertaining to the electronic structure
and excitation spectra of atoms,
molecules and condensed phases. Materials
clusters exhibit a high sensitivity 
of the electronic spectrum to the
geometrical structure which often 
differs from that of the bulk,
and show a high propensity  to form
structural isomers  
that may dynamically interconvert  at
finite temperatures. Consequently,
high-resolution PES emerges as an important
tool in cluster science, particularly 
in the face of severe difficulties
in applying common direct structure-determination
techniques to such systems.

However, a reliable interpretation of PES
spectra is often theoretically challenging
due to several factors, including:
finite-state effects, electronic and ionic
relaxations, thermal ionic motions,
and structural isomerizations.
With the advent of accurate ab-initio
methods for electronic structure
calculations, theoretical investigations
of some of these issues have been pursued
\cite{koutecky,cheli,car}.
Particularly pertinent to our study is
the development of methods which allow
practical and reliable simulations
of PES spectra including dynamical
(finite-temperature) effects [2-4a].

In this paper we address, via the use
of ab-initio BO-LSD-MD (Born-Oppenheimer
local-spin-density molecular dynamics)
simulations [4a], 
methodological issues pertaining to simulations
and analysis of finite-temperature 
PES spectra. We performed accurate
(and practical) calculations of PES
spectra from recorded density of states of the
clusters  using  a
"generalized Koopmans' theorem" (GKT)\cite{Tozer},
concurrent with  simulations of the ionic dynamics.
Furthermore, in conjunction with
 measured \cite{Wang1}
high-resolution PES spectra for
Al$_n^-$ (12$\leq$n$\leq$15) cluster anions,
we illustrate that the simulated spectra
provide a (quantitatively) faithful
description of the measured ones,
including thermal  effects,
thus allowing reliable assignments 
of ground as well as isomeric structures.
Additionally, we demonstrate 
that through comparisons between simulated 
PES spectra and those measured in three
(experimentally undetermined) temperature
regimes estimates of the clusters'
temperatures can be obtained.

 In the BO-LSD-MD 
 method the motions of the ions evolve in time
according to classical equations of motion,
with the electronic 
  Hellmann-Feynman
forces evaluated  at each MD time step 
via a self-consistent solution
of the Kohn-Sham (KS) equations, 
using the LSD exchange-correlations  after Ref. 
\cite{Vosko}, and in conjunction with non-local norm-conserving  
 pseudopotentials \cite{Troullier}.    
An important element of the method,
distinguishing it from those used previously
in PES studies \cite{cheli,car},  is the
fact that it does not employ supercells
(periodic replicas of the system), and consequently 
charged systems as well as those having permanent
and/or dynamically developed multipole moments
are simulated accurately in a straightforward manner
on equal footing with neutral ones (i.e. alleviating
the need for an artificial neutralizing
background, large calculational cells, and/or
approximate treatment of long-range multipole
image interactions).

The ground state structures  of Al$_{12}^-$--Al$_{15}^-$,
determined by us through structural optimization
starting from those of the corresponding neutral
clusters [4b], are displayed in Figure 1.
Aluminum clusters in this size range 
favor energetically  icosahedral-based structures \cite{Uzi1};
Al$_{12}^-$ having an oblate 
deformed shape,
that of Al$_{13}^-$ being close to an
ideal icosahedron,  and those of 
Al$_{14}^-$ 
and Al$_{15}^-$ being capped icosahedra.
For Al$_{15}^-$  we find that 
 in the energy-optimal
structure the two capping atoms are located on the opposite
sides of a "core" icosahedron, resulting in a 
strongly deformed prolate shape (see Fig. 1) \cite{Uzi2}.

%%%% FIGURE 1 %%%%%%%%%%%%

The electronic structure of the ground state
cluster anions exhibits sizable 
gaps (E$_g$) between the highest-occupied
KS molecular orbitals (HOMO) and the lowest
unoccupied ones (LUMO), as well as odd-even
alternation (as a function of the number of
electrons) 
in the vertical
detachment energy (vDE) shown in Table I. 
Al$_{13}^-$ is electronically
"magic" (i.e. 40 valence electrons), having 
an exceptionally high vDE 
\cite{Wang1} and the largest E$_g$. 
Its electronic structure reflects 
the corresponding neutral cluster, 
which was found 
[4b] to exhibit  
a clear jellium-type filling sequence
$1s^21p^61d^{10}2s^2$ for
the lowest 20 single-particle states.
The remaining 20 states, which would
correspond to jellium $1f^{14}2p^6$ states,
 are grouped
into two broadly overlapping subbands
(finite temperature broadening of these
bands is displayed in Figure 2)
and show significant $pf$ mixing;
this level scheme is known to be a consequence  
of the $I_h$ icosahedral  symmetry.

%%%%%%%%%%%%%% TABLE 1 %%%%%%%%%%%%%%%%%%%%%%%%%

Although the KS states are not necessarily
the "true" molecular orbitals of the system,
it has been observed that the KS HOMO eigenvalue
of the  $N$-electron system,
 $\epsilon_{HOMO}(N)$, bears a well-defined
relation to the ionization potential $I(N)$ and
electron affinity $A(N)$ \cite{Perdew,Tozer};
through Janak's theorem \cite{Janak}.  
Following these ideas we make use here 
of a   "generalized Koopmans' theorem" (GKT) \cite{Tozer,Note1}  
\begin {equation}
\epsilon_{HOMO}(N)-v_{xc}^{\infty}=-I_{GKT}(N),
\end {equation}
where  $v_{xc}^{\infty}$ 
is the asymptotic limit of the exchange-correlation 
potential. This nonvanishing energy shift 
is required for an accurate description of the
asymptotic KS equations \cite{Tozer}.
While rigorously the vertical detachment energy 
would be given by $E(N-1)-E(N)$, where $E(N)$ and
$E(N-1)$ are, respectively, the total energies of
the anion and neutral (unrelaxed) clusters,
Eq. (1) suggests a practical approach for evaluation of the 
threshold region of finite-temperature PES spectra through MD
simulations.
Accordingly, neglecting hole-relaxation effects, the vDE 
for removing the electron from the HOMO state 
may be well estimated   by  ($-\epsilon_{HOMO}+v_{xc}^{\infty}$),
 provided that  
 $v_{xc}^{\infty}$ remains constant to a 
good approximation, regardless of  
 spatial details of the $N$--electron
system (such as isomeric atomic configurations
of the cluster). To explore the validity of
this condition 
we  have calculated for each of the
cluster anions the energy shift 
$v_{xc}^{\infty}=E(N-1)-E(N)+\epsilon_{HOMO}(N)$
for a selected  set of structures (including
the ground state one  and 10 other
configurations chosen randomly  from finite-temperature
MD trajectories). 
The calculated values of
$v_{xc}^{\infty}$ were found to have a spread of 
no more than   0.04 eV
for each of the clusters, and  furthermore we
found that the dependence of
 $v_{xc}^{\infty}$ 
on $N$ ($37\leq N\leq 46$) is very weak (see Table I).

While this procedure could be repeated 
to determine $v_{xc}^{\infty}$ values
for vDEs from deeper (lower-energy)
KS states,
we have chosen to use a simpler (and more 
practical) procedure whereby
we use the shift calculated for the HOMO
level also for the deeper states \cite{Note2}.
In this way we generate the full PES spectra
from the density of states (DOS) recorded 
in the course of the BO-LSD-MD simulations \cite{Note3}.
As shown below, this theoretically founded procedure
yields a 
faithful description of the experimental
data, and it is a viable and reliable
alternative to previously used methods
for finite-temperature ab-initio MD
simulations of PES spectra which were based
on either ad-hoc shifts (aligning the theoretical
DOS with the dominant features in the
measured spectra) \cite{cheli} or
on a first-order pertubative treatment [3b].
Furthermore, the comparative analysis 
of the simulated PES spectra and the measured
ones (see below) validates a posteriori
certain general assumptions underlying
MD simulations of PES spectra, i.e.:
neglect of finite-lifetime effects of
the hole  (see also Ref. \cite{cheli}
where it is noted that such effects 
may contribute only for very  small
clusters);
the use of vDEs (i.e. neglect of
ionic relaxations following the
detachment process); 
 and, assumed equal weights for all states
contributing to the PES spectrum (i.e. 
neglect of photoelectron transition-matrix
effects, which may affect  line-shape
features and certainly the absolute 
PE cross-sections, but  not 
the locations of spectral features
i.e. binding energies).

The measured  PES spectra for Al$_{12}^-$--Al$_{15}^-$ 
are shown in 
 Figures 2 and 3 (solid line). 
It is found 
 that clusters leaving the nozzle 
early (short residence time) are quite "hot"
 whereas clusters leaving the nozzle late 
(long residence time) are "colder".
Indeed the PES spectra for the cold
clusters shown in Fig. 2  and the bottom panel of Fig. 3
 exhibit well-defined features.
On the other hand,
 hot clusters 
exhibit  much broader and diffused spectral features,
 as shown
 in Figure 3 for Al$_{13}^-$, where we display spectra 
measured for  
 three different residence times,  labeled as
"cold", "warm", and "hot".  
Comparisons between the locations (binding energies)
of the peaks and shoulders in the measured and
simulated spectra for the cold clusters (simulation
temperatures of 130 K to 260 K, see
 caption to Fig. 2),
validate the  $v_{xc}^{\infty}$--shifting
procedure of the calculated DOS described above.
The widths of the peaks in the theoretical 
PES spectra originate solely from atomic
thermal vibrations since at these 
low temperatures isomerization effects
and/or strong shape fluctuations do not occur.
The good agreement achieved here, without
any adjustable parameters other than the
ionic temperature in the MD simulations,
strongly indicates that the "cold" clusters
in the experiments  are indeed well
below room temperature.

%%%%%%%% FIGURE 2 %%%%%%%%%%%%%%%

Theoretical PES spectra 
corresponding to isomeric structures of
Al$_{13}^-$--Al$_{15}^-$, calculated at 0 K,
are also shown in Fig. 2 (see inset
for the threshold regions of Al$_{13}^-$,
 and the dotted line
 in the  panels for Al$_{14}^-$ and Al$_{15}^-$).
The isomers for Al$_{13}^-$ and Al$_{14}^-$
are the aforementioned decahedral ones \cite{Uzi1}, and
in the Al$_{15}^-$ isomer two neighboring
triangular facets are capped. 
Comparison between these spectra and those 
calculated for the ground state clusters 
as well as with the measured ones,
suggests overall that at low temperatures
either these isomers  do not
occur, or that their abundance in the cluster beam
is rather low. In this context we note that starting
from the decahedral isomer of Al$_{13}^-$,
it transformed readily during  
short MD simulations into the icosahedral one 
at about room temperature.
 This supports our conclusion
pertaining to the low abundance in the cold beam
of clusters
"trapped" in isomeric structures;
however, an even small relative (quenched) 
concentration of such isomer in the cold 
Al$_{13}^-$ beam may be sufficient to account
for the low-binding energy tail observed in the measured 
PES spectra for Al$_{13}^-$ (see inset in Fig. 2).

%%%%% FIGURE 3 %%%%%%%%%%%%%%%%%%%%

Both the experimental and theoretical PES spectra,
shown in Figure 3 for Al$_{13}^-$, which
were measured at the three temperature regimes
mentioned above and simulated at the indicated
temperatures, exhibit gradual broadening
and "smearing" of the PES spectral
features as the temperature increases. We also
observe that the binding energy of the main
peak is rather insensitive to the thermal
conditions, while the line-shape near the
threshold region (lower binding energies)
exhibits a rather pronounced temperature
dependence.

The broadening of the spectral features 
and the (so called) "thermal tail effect"
near threshold originate from the 
variations of the electronic structure
caused by enhanced vibrational motions at
the higher temperatures, as well as from increased
isomerization rates (e.g. in the "warm"
regime) governed by the free-energy
of the cluster (that is enhanced contributions
of lower frequency modes to the
vibrational entropy \cite{Martin}),
and from disordering ("melting")
of the cluster in the "hot" regime
(where inspection of the atomic trajectories 
 reveals frequent
transitions between a broad assortment
of  configurations).
Indeed, examination of the 
vibrational DOS of the simulated clusters
(obtained via Fourier transformation of the
atomic velocity autocorrelation
functions) revealed a marked gradual
softening of the clusters at the
"warm" and "hot" regimes (that is
shifting of the vibrational spectrum
to lower frequencies) coupled with increasing
overlap between the frequency regions
of the various modes due
to large anharmonicities.

In light of the above we judge
the overall agreement between the simulated
and measured spectra and their 
thermal evolution as rather satisfactory,
and the remaining discrepancies 
(mainly in line-shapes) 
 may be
attributed to insufficient sampling
during the 5 ps MD
simulations of the thermally-expanded
phase-space of the clusters.
 
The methodology developed in this study for
practical calculations of finite-temperature
PES spectra, through ab-initio MD simulations
of aluminum cluster anions 
 with no adjustable
parameters other than the 
clusters' temperatures,
was demonstrated to yield results in
agreement with high-resolution 
PES spectra measured at various
thermal conditions of the cluster beam.
Such comparative analysis
allows reliable structural
assignments and theoretical estimation
of the clusters' temperatures,
as well as gaining  insights
into the electronic and structural
properties of clusters and their
thermal evolution.

\bigskip

Computations were performed mainly
on Cray T3E at the Center for Scientific Computing,
Espoo, Finland, and in part 
on an IBM RISC 6000/SP2
at the GIT Center for Computational Materials Science. 
Work in the University of Jyv\"askyl\"a is
supported  
by the Academy of Finland, and at Georgia Tech
by the US DOE. 
The experimental work is supported by the NSF
and performed
at EMSL, a DOE user facility located at PNNL,
which is operated for DOE by Battelle
Memorial Institute. 
L. S. W. acknowledges support from the Alfred P. Sloan Foundation.
J. A. wishes to thank the V\"ais\"al\"a Foundation
for support.

%%%%%%%%%%%%%% TABLE 1 %%%%%%%%%%%%%%%%%%%%%%%%%
\begin{table}
\caption{Number of valence electrons ($N$) in 
the cluster anions, HOMO-LUMO 
gap (E$_g$), 
 vertical detachment energy (vDE) of 0 K  ground
state anion, and numerically determined asymptotic
exchange-correlation shift ($v_{xc}^{\infty}$). 
Energies in eV.}

\label{tab1}
\begin{tabular}{rrrrr}
  & $N$  & E$_g$ & vDE & $v_{xc}^{\infty}$ \\
\tableline
Al$_{12}^-$ & 37 & 0.25 & 2.82 & 1.63 \\
Al$_{13}^-$ & 40 & 1.89 & 3.59 & 1.64 \\
Al$_{14}^-$ & 43 & 0.32 & 2.67 & 1.57 \\
Al$_{15}^-$ & 46 & 0.79 & 3.07 & 1.57 \\
\end{tabular}
\end{table}
%%%%%%%%%%%%%%%%%%%%%%%%%%%%%%%%%%%%%%%%%%%%%%%%

%%%% FIGURE 1 %%%%%%%%%%%%
\begin{figure}
\caption{
The ground state geometries of Al$_{12}^-$-Al$_{15}^-$
(left to right).
}
\end{figure}
%%&&&&&&&&&&&&&&&&&&&&&&&

%%%%%%%% FIGURE 2 %%%%%%%%%%%%%%%
\begin{figure}
\caption{
Measured     photoelectron spectra
of cold (long residence time) Al$_{12}^-$- Al$_{15}^-$
at 193 nm (solid lines) compared to the
simulated 
 spectra (dashed lines). The simulation 
temperatures are 160, 260, 200, and 130 K for
 Al$_{12}^-$--Al$_{15}^-$, respectively.
The arrows correspond to the vDE of the ground
state structure at 0 K, given in Table I.
The inset shows the 0 K PES spectra
for the ground state (solid line) and 
the decahedral isomer (dashed line)
of Al$_{13}^-$. 
The dotted line for Al$_{14}^-$ is
the 0 K  spectrum of the decahedral 
isomer, and that for Al$_{15}^-$ is the
0 K spectrum of an icosahedral-based
isomer (see text). 
}
\end{figure}
%%%%%%%%%%%%%%%%%%%%%%%%%%%%%%%%%%%%%

%%%%% FIGURE 3 %%%%%%%%%%%%%%%%%%%%
\begin{figure}
\caption{
Measured
temperature-dependent PES spectra of Al$_{13}^-$
(solid lines) compared to the simulated ones 
at 930 K, 570 K
and 260 K (dashed lines). HOT - short residence time,
WARM - medium residence time, COLD - long residence time.
}
\end{figure}
%%%%%%%%%%%%%%%%%%%%%%%%%%%%%%%%%%%

\end{multicols}

\begin{thebibliography}{99}


\bibitem{koutecky}
V. Bonacic-Koutecky {\it et al.}, 
J. Chem. Phys. \textbf{93}, 3802 (1990);
{\it ibid.}, \textbf{100}, 490 (1994).

\bibitem{cheli}
N. Binggeli and J. R. Chelikowsky, Phys. Rev. Lett. \textbf{75}, 493 (1995).

\bibitem{car}
(a) C. Massobrio {\it et al.}, Phys. Rev. Lett. \textbf{75},
2104 (1995); (b) Phys. Rev. B \textbf{54}, 8913 (1996).

\bibitem{Barnett}
(a) For a formulation and details of the 
BO-LSD-MD method see:
 R.N. Barnett and U. Landman, Phys. Rev. B \textbf{48}, 2081 (1993).
(b)
For  recent studies of Al clusters using this
method  see:
J. Akola {\it et al.}, Phys. Rev. B \textbf{58}, 
3601 (1998);  Europ. Phys. J D (Vol. 9, in print (1999)).


\bibitem{Tozer}
D.J. Tozer and N.C. Handy, J. Chem. Phys.  
\textbf{108}, 2545 (1998); {\it ibid.},  \textbf{109}, 10180 (1998).

\bibitem{Wang1}
X. Li {\it et al.},
Phys. Rev Lett. \textbf{81}, 1909 (1998).
The experimental setup described in this paper
is the same as that used in the 
current study, where in addition, here
PES spectra were measured  also at various
temperature regimes (see discussion in 
connection with Figs. 2 and 3).            

\bibitem{Vosko}
S.H. Vosko {\it et al.},
Can. J. Phys. \textbf{58}, 1200 (1980);
S.H. Vosko and L. Wilk, J. Phys. C \textbf{15}, 2139 (1982).

\bibitem{Troullier}
N. Troullier and J. L. Martins, Phys. Rev. B \textbf{43}, 1993 (1991).
For the aluminum $3s^23p^1$ valence electrons, we use 
$s$-nonlocal and $p$-local components with
cut-off radii of 2.1 and 2.5 $a_0$, respectively.
The KS orbitals are expanded in a  plane wave basis
with a  cutoff of 15.4 Ry. 

\bibitem{Uzi1}
In Ref. [4b] it has been noted that a
truncated decahedral motif, obtained from
the icosahedron by a $\pi/5$ rotation
of the two opposing capped pentagons with 
respect to each other, is energetically close
to the icosahedral geometry of
Al$_{13}$ -- Al$_{15}$. On the other
hand, fcc-based cuboctahedral structure
are much higher in energy in this
size range.
For other recent ab-initio studies
of aluminum clusters in this
size range, see references in [4b].


\bibitem{Uzi2}
The pattern of shape
deformations for all the cluster anions shown in Fig. 1
correlates well with that obtained via
jellium calculations for clusters with the
number of electrons corresponding to that in
this size range (37 to 46 electrons).
 For a review, see C. Yannouleas and
U. Landman, in "Large Clusters of Atoms
and Molecules",  T. P. Martin, Ed.
(Kluwer, Dordrecht, 1996), p. 131.


\bibitem{Perdew}
J.P. Perdew {\it et al.},
Phys. Rev. Lett. \textbf{49}, 1691 (1982);
J.P. Perdew and M. Levy,  Phys. Rev. Lett. \textbf{51}, 1884 (1983);
J.P. Perdew and K. Burke, Int. J. Quantum. Chem. \textbf{57}, 309 (1996).

\bibitem{Janak}
J. F. Janak, Phys. Rev. B \textbf{18}, 7165 (1978).

\bibitem{Note1}
Koopmans' theorem, $\epsilon_{HOMO}(N)=-I(N)$, 
originally proposed in the context of
the  Hartree-Fock method,
provides often a reasonable estimate for
$I$ (within the method),
 and is in a wide use in quantum chemistry.

\bibitem{Note2} 
Restricted (spherical)
jellium calculations ($r_s$(Al)=2.07 $a_0$)
for a 40 electron system (Al$_{13}^-$),
where electrons
are removed from deeper levels  (with a "frozen-orbital"
approximation), 
show that the $v_{xc}^{\infty}$ shifts
for these deeper levels are approximately
constant to within 0.2 eV
 [M. Manninen, unpublished].

\bibitem{Note3}
Microcanonical  MD simulations were performed at various
temperatures around and below room temperature,
and Al$_{13}^-$ was also simulated 
at 570 K and 930 K. 
A 5 fs time step 
was used in integration of the classical equations
of motion, and  the DOS data was recorded  
in  5 ps simulations (after equilibrating the
system at the desired temperature for    
about 2 ps). The DOS spectra were 
smoothed by replacing every data point
by a 0.05 eV FWHM Gaussian. These spectra were
then shifted by the $v_{xc}^{\infty}$ values given
in Table I for each cluster anion.  
This gives the absolute binding energy scale
on which the calculated spectra are plotted
in Figures 2 and 3. 
The height of the calculated spectra are
scaled to the main peak of the experimental
spectra in Al$_{13}^-$ to Al$_{15}^-$ and 
to the second peak in the case of Al$_{12}^-$.

\bibitem{Martin}
T. P. Martin, Phys. Rep. \textbf{95}, 167 (1983).


\end{thebibliography}
\end{document}